\documentclass[twocolumn,preprintnumbers,amsmath,amssymb]{revtex4}  

\usepackage{graphicx}
\usepackage{dcolumn}
\usepackage{bm}
\usepackage{multirow}
\usepackage{color}
\usepackage{amsmath}

\begin{document}
\title{Double resonant Raman scattering and valley coherence generation\\ in monolayer WSe$_2$}

\author{G.~Wang$^1$}
\author{M.~M.~Glazov$^{2,1}$}
\author{C.~Robert$^1$}
\author{T.~Amand$^1$}
\author{X.~Marie$^1$}
\author{B.~Urbaszek$^1$}
\affiliation{%
$^1$ Universit\'e de Toulouse, INSA-CNRS-UPS, LPCNO, 135 Av. Rangueil, 31077 Toulouse, France}
\affiliation{%
$^2$ Ioffe Institute, 26 Polytechnicheskaya, 194021 St.-Petersburg, Russia}


\begin{abstract}
The electronic states at the direct band gap of monolayer WSe$_2$ at the $K^+$ and $K^-$ valleys are related by time reversal and may be viewed as pseudo-spins. The corresponding optical interband transitions are governed by robust excitons. In double resonant Raman spectroscopy, we uncover that the $2s$ exciton state energy differs from $1s$ state energy by exactly the energy of the combination of several prominent phonons. Superimposed on the exciton photoluminescence (PL) we observe the double resonant Raman signal. This spectrally narrow peak shifts with the excitation laser energy as incoming photons match the $2s$ and outgoing photons the $1s$ exciton transition. The multi-phonon resonance has important consequences: Following linearly polarized excitation of the $2s$ exciton a superposition of valley states is generated which can relax fast via phonon emission and with minimal loss of coherence from the $2s$ to $1s$ state. This explains the high degree of valley coherence measured for the $1s$ exciton PL. 
\end{abstract}


                             
\maketitle
\emph{Introduction.---} Monolayer (ML) transition metal dichalcogenides (TMDCs) such as MoS$_2$ and WSe$_2$ have a direct optical bandgap at the $K$-points of the Brillouin zone \cite{Mak:2010a,Splendiani:2010a,Zhao:2013b}. Light absorption and emission are governed by robust excitons, Coulomb bound electron-hole pairs, as their binding energy and oscillator strength is greatly enhanced in these ideal two-dimensional (2D) systems. The electronic states in the $K^+$ and $K^-$ valleys are related by time reversal and may be viewed as pseudo-spins. The electrons can be optically initialized either in the $K^+$ or $K^-$ valley using the chiral optical selection rules, i.e. with circularly right ($\sigma^+$) or  left ($\sigma^-$) polarized light, respectively  \cite{Xiao:2012a, Cao:2012a}. Hence, the valley degree of freedom can be manipulated by light~\cite{Mak:2014a,Xu:2014a}, an opportunity to investigate Berry phase effects in solid-state physics \cite{Xiao:2010a}.
\begin{figure}[t]
\includegraphics[width=0.48\textwidth]{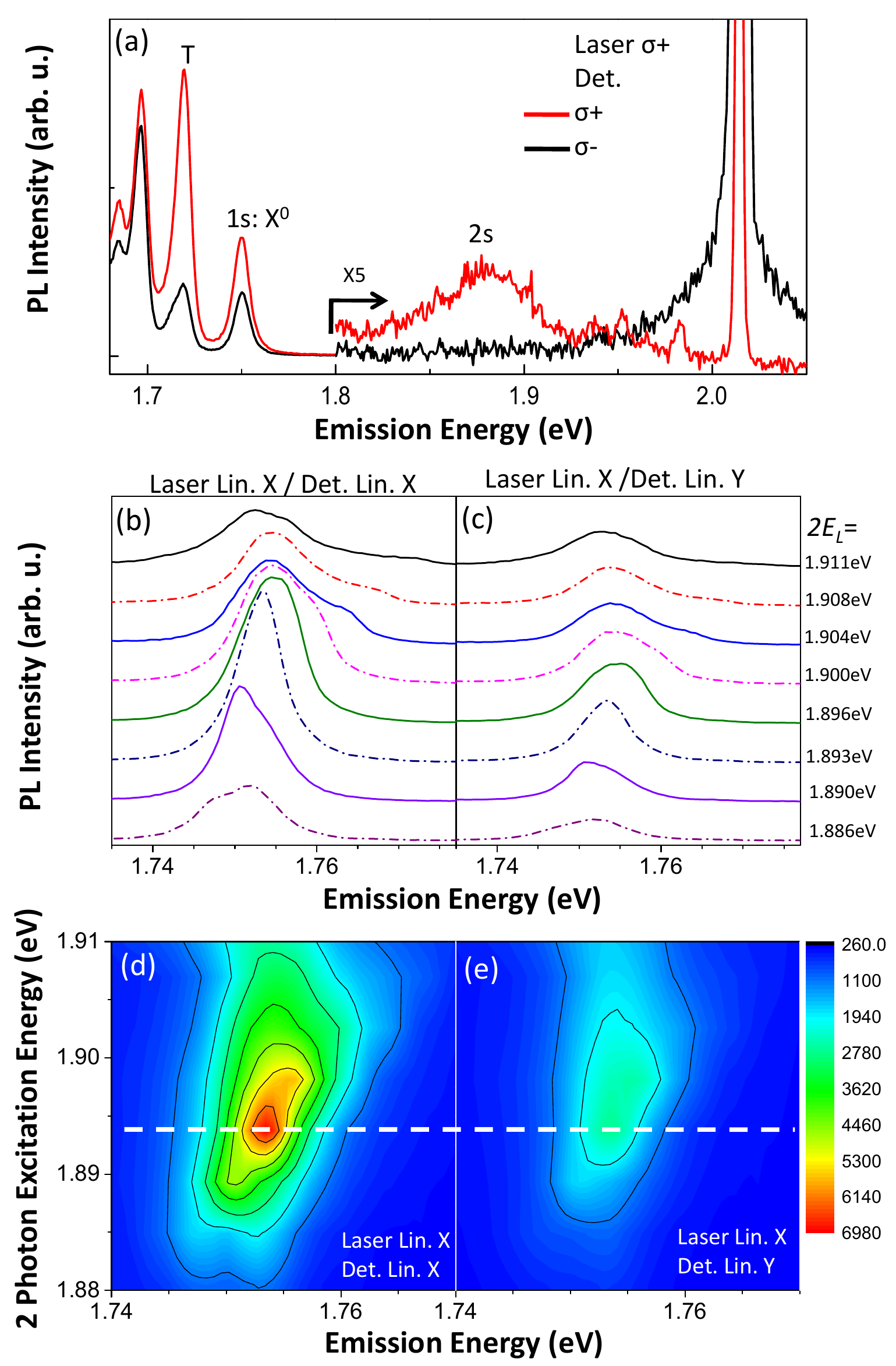}
\caption{\label{fig:fig1} \textbf{2-photon excitation (a)} Broad $2s$ exciton PL peak following excitation of $3p$ state. Trion (T) and neutral exciton X$^0$ (1s) are marked. \textbf{(b)} emission at the X$^0$ energy ($1s$ exciton) for different laser energies $2E_L$ close to the $2p$ exciton resonance. Laser polarization linear X, detection linear X (co-polarized). \textbf{(c)} same as (b) but for detection  linear Y (cross-polarized). \textbf{(d)} same data as (b) but presented as a contour plot to show changes in PL intensity and shape as a function of $2E_L$, dashed line marks $2p$ resonance. \textbf{(e)} same data as (c), same colour scale as (d). 
}
\end{figure}
Highly circularly polarized photoluminescence (PL) reported for WSe$_2$ \cite{Jones:2013a,Wang:2015b}, MoS$_2$ \cite{Mak:2012a,Zeng:2012a,Kioseoglou:2012a,Sallen:2012a} and WS$_2$~\cite{Zhu:2014b} confirms efficient and robust optical valley polarization initialization. For coherent manipulation of arbitrary valley states, efficient interaction with linearly and eventually elliptically polarized laser light needs to be demonstrated. Excitation with linearly polarized light yields the coherent superposition of exciton pseudo-spin states, a process called optical alignment of excitons \cite{Meier:1984a}. As each exciton state in ML TMDCs is built from electrons and holes from the  $K^+$ and $K^-$ valleys, optically aligned exciton states are also referred to as coherent valley states \cite{Jones:2013a}. However, the process of optical generation of valley coherence and its detection in linearly polarized PL emission has been more elusive~\cite{Jones:2013a,Zhu:2014b,Wang:2015b}. \\
\indent In most semiconductor systems (pseudo-)spin coherence is lost during energy relaxation involving scattering events. As a  consequence strictly resonant laser excitation is favoured for coherent control schemes \cite{Dyakonov:2008a,Greilich:2009a,Marie:1997a}. In this work we demonstrate how and why valley coherence can be generated efficiently capitalizing on a multi-phonon resonance that we uncover in double resonant Raman scattering \cite{Miller:1986a,Cerdeira:1986a,Ferrari:2006a,Thomsen:2000a,Maultzsch:2003a}. We combine two remarkable properties of ML TMDCs in our experiments, namely (i) the dominant role of excitons \cite{Cheiwchanchamnangij:2012a,Komsa:2012a,Song:2013a,He:2014a,Ugeda:2014a,Chernikov:2014a,Ye:2014a,Wang:2015b,Klots:2014a,Zhu:2015b,Hanbicki:2015a} combined with (ii) the strong interaction of excitons with phonons \cite{Carvalho:2015a,Zhang:2015a,Fan:2014a,Delcorro:2014a,Gaur:2015a}.  \\
\begin{figure*}[ht!]
\includegraphics[width=0.99\textwidth]{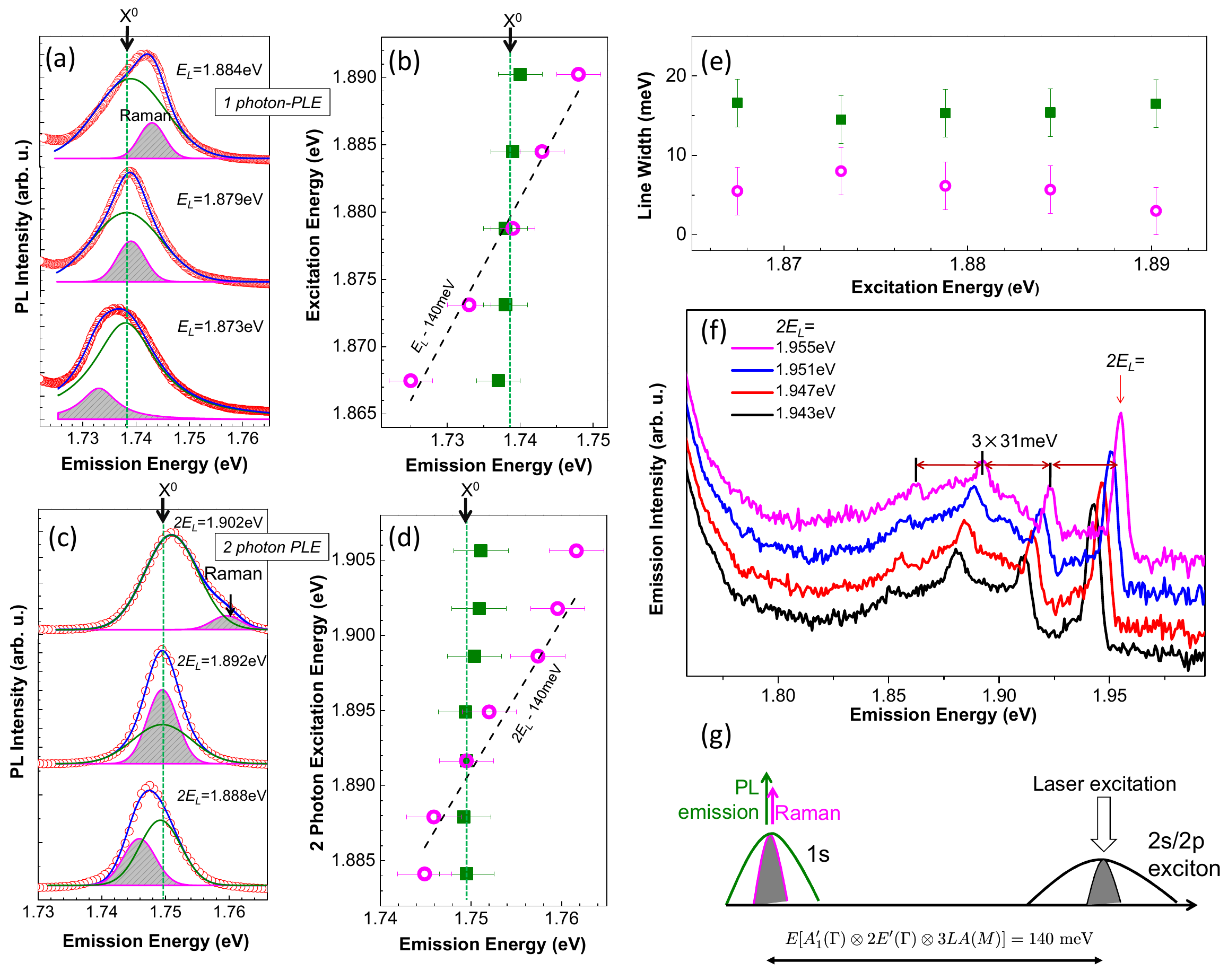}
\caption{\label{fig:fig2}  1ML WSe$_2$, T=4~K. \textbf{(a)} emission at exciton X$^0$ 1s energy (red circles) for different excitation laser energies $E_L$ for sample 1, spectra are offset for clarity. Two contributions can be separated via fitting: The X$^0$ PL component (green) and the resonant Raman contribution (pink, shaded grey), the latter changes with $E_L$ from top to bottom panel. \textbf{(b)} Extracted transition energies for X$^0$ PL (green squares) are constant, resonant Raman spectral position (open circles) changes linearly as a function of laser excitation energy, for comparison we plot $E_L-140$~meV (dashed line). \textbf{(c)} and \textbf{(d)} same as (a) and (b) but using 2-photon excitation for sample 2. \textbf{(e)}  Line width of the X$^0$ PL (green squares) and resonant Raman signal (pink circles) as a function of laser excitation energy as extracted from panel (a). \textbf{(f)} Simply resonant Raman scattering as outgoing photons are in resonance with the broad 2s exciton state shows 3 phonon replica more than 90~meV below the 2-photon excitation. \textbf{(g)} schematics of exciton generation and relaxation via multiple phonon processes, leading to enhanced X$^0$ emission and a double resonance (incoming and outgoing photons) for Raman scattering.
}
\end{figure*} 
\indent We show that valley coherence generation in ML WSe$_2$ is most efficient when the excitation laser is exactly 140~meV above the $1s$ exciton. This has two reasons: First, light absorption is enhanced as states of the $2s$ (or $2p$) exciton are available. Second, fast and efficient energy relaxation with minimal coherence loss to the $1s$ exciton occurs via multi-phonon emission. We observe that the exciton PL emission from coherent valley states is superimposed on a strong double resonant Raman scattering signal (not linked to valley physics) as the incoming and outgoing photon energies are in resonance with the $2s$ and $1s$ exciton states, respectively.
We are able to spectrally separate the exciton emission and the spectrally narrower Raman signal by systematically varying the laser energy. We show that the Raman signal provides a substantial contribution to the global emission and its linear polarization. Therefore care needs to be taken when analyzing strongly linearly polarized emission from the $1s$ exciton. Double resonances are much stronger than single resonances \cite{suppl}, making double resonant Raman scattering a sensitive tool for investigating the electronic transitions in, for example, graphene, graphite and carbon nanotubes \cite{Ferrari:2006a,Thomsen:2000a,Maultzsch:2003a}. Thus, the observation of double resonant Raman scattering mediated by excitons  in TMDCs opens up possibilities for in-depth studies of fine details of the energy spectrum, currently debated in the literature \cite{Cheiwchanchamnangij:2012a,Komsa:2012a,Song:2013a,He:2014a,Ugeda:2014a,Chernikov:2014a,Ye:2014a,Wang:2015b,Klots:2014a,Zhu:2015b,Hanbicki:2015a}.\\
\indent \textit{Samples and Experimental Set-up.---}
The WSe$_2$ ML flakes are obtained by micro-mechanical cleavage of a bulk crystal on SiO$_2$/Si substrates. Experiments at $T=4$~K are carried out in a confocal microscope \cite{Wang:2014b}. The detection spot diameter is $\approx1\mu$m, i.e. considerably smaller than the ML size of $\sim 10~\mu$m$\times10~\mu$m. For time integrated experiments, the PL emission is dispersed in a spectrometer and detected with a Si-CCD camera. Our target is to distinguish between PL emission and spectrally narrow Raman signals. The WSe$_2$ ML is therefore excited by \textit{picosecond} pulses generated by a tunable frequency-doubled optical parametric oscillator (OPO) synchronously pumped by a mode-locked Ti:Sa laser. The typical pulse temporal and spectral widths are 1.6 ps and 3 meV, respectively, the latter being smaller than the exciton PL emission linewidth 
$\sim 15$~meV; the repetition rate is 80 MHz.  \\
\indent \textit{Results and Discussion.---}
In our measurements with a tunable laser source, we change the excitation energy and monitor the PL emission of the WSe$_2$ ML. We concentrate here on the emission of the neutral exciton X$^0$ ground state ($1s$) at 1.75~eV, with a FWHM of about $\sim 15$~meV, see Fig.\ref{fig:fig1}a. 
At 1.72~eV we observe the charged exciton emission (trion, T). The emission observed at lower energies is probably related to localized states, it disappears when the temperature is raised \cite{Wang:2014b}. 
In previous work \cite{Wang:2015b}, we have identified the position of the $2s$ and the $2p$ exciton states in 1-photon and 2-photon PL excitation (PLE) experiments, respectively. Interestingly, in ML WSe$_2$, both states are very close in energy (around 1.89 eV). Here we show in addition hot luminescence of the $2s$ state in Fig.~\ref{fig:fig1}a as a result of resonant 2-photon excitation of the $3p$ state at about 2.03~eV. Crucially for the experiments presented here, resonant excitation of the excited ($2s/2p$) exciton state results in a strong enhancement of the $1s$ exciton PL. Importantly, using a linearly polarized laser to excite the $2s/2p$ state, the resulting $1s$ emission is strongly linearly polarized $\sim 40\%$ (compare intensities in Fig.\ref{fig:fig1}d vs.  Fig.\ref{fig:fig1}e and data for 1-photon PLE in the supplement \cite{suppl}), a clear fingerprint of efficient valley coherence generation i.e. exciton alignment~\cite{Jones:2013a}. It is very surprising to observe the recombination of a coherent superposition of valley states using optical excitation as high as 140~meV above the $1s$ state. \\
\indent Our 2-photon PLE experiments shown in Figs.\ref{fig:fig1}b-e clearly demonstrate that the variation of the exciton laser energy has several effects on photons emitted at the $1s$ energy: (i) excitation close to the  $2p$ exciton results in drastic changes of the X$^0$ PL emission intensity (ii) as the laser is scanned through the $2p$ resonance, the line shape and central energy of the emission are modified. Equivalent results for 1-photon PLE are shown in the supplement \cite{suppl}. A detailed analysis of the shape and energy position of the X$^0$ emission for different laser energies is shown in Fig.~\ref{fig:fig2}a for 1-photon PLE. When exciting close to the $2s$ resonance, the emission consists of \emph{two} distinct features: In addition to the strong and highly polarized X$^0$ emission (energy position confirmed for more than 50 values of non-resonant laser excitation energies) we observe a spectrally much narrower feature, that changes energy as the laser energy is changed. Careful fitting of the emission for different laser energies confirms that the additional, narrow feature is observed exactly 140~meV below the laser excitation energy, as shown clearly in Fig.~\ref{fig:fig2}b. In Fig.~\ref{fig:fig2}e we show that the linewidth of the X$^0$ PL emission is about 15~meV and the narrower contribution has a FWHM of about 5~meV \cite{phononbroad}. We have confirmed this observation in several samples. In Fig.~ \ref{fig:fig2}c we show additional evidence for a different sample:  Using 2-photon excitation, we also find a strong resonance when twice the laser energy $2E_L$ is 140~meV above the $1s$ exciton. Again we analyse the X$^0$ emission around this resonance: Two features, one moving with the laser energy the other one not, can be discerned. Fitting the results in Fig.~\ref{fig:fig2}c clearly reveals a narrow feature that is always 140~meV below $2E_L$, in addition to the X$^0$ PL with fixed  emission energy, see Fig.~\ref{fig:fig2}d.\\
\indent Our experimental results raise the following questions: What is the physical origin of the spectrally narrow peak whose energy is fixed with respect to that of the laser? Why is the X$^0$ coherence and emission intensity at a maximum when this narrow peak centre energy is exactly at the X$^0$ $1s$ resonance? A maximum in the PLE signal requires (a) strong absorption by an excited state, here the $2s$ or $2p$ exciton, and (b) efficient relaxation from the excited state to the exciton ground state ($1s$). Relaxation between electronic states in semiconductors is most efficient by emission of phonons. This allows in addition to partially preserve coherence \cite{Flissikowski:2001a}. Raman studies in ML TMDC have revealed a multitude of well characterized phonon modes, e.g. see \cite{Carvalho:2015a,Zhang:2015a,Fan:2014a,Delcorro:2014a,Jimenez:1991a,Ataca:2011a,Luo:2013a,Sahin:2013a,Zhao:2013a}. 
A key point in our experiment is the enhancement of light scattering by phonons as the photon energy is in resonance with an electronic transition. In Fig.~\ref{fig:fig2}f we present an example of resonant Raman scattering in our sample using a single (not double) resonance, similar to recent experiments in ML MoS$_2$ \cite{Carvalho:2015a}: Excitons are generated via 2-photon absorption, and up to three optical phonons (with energy~31~meV~\cite{note:3phonons}) are emitted, so that excitonic final states fall into the broad line of the $2s/2p$ state. This simple resonant Raman experiment shows \emph{three} clear phonon replica more than 90~meV below the excitation energy. Multiple Raman features are often observed in 2D materials such as GaSe \cite{Reydellet:1976a} and TMDCs \cite{Zhang:2015a} as the interaction of phonons with exciton transitions is efficient. Due to 2D confinement, exciton oscillator strength and binding energy are enhanced and emission linewidths in the meV range allow to see clear resonance effects. For $2s/2p$ to $1s$ phonon-assisted transitions investigated here in ML WSe$_2$ an ideal situation for \textit{double} resonant Raman scattering occurs: The incoming photons are in resonance with the $2s/2p$ exciton state and the outgoing photons with the $1s$ exciton state, see Fig.~\ref{fig:fig2}g. Double resonant Raman scattering is extremely efficient as the scattering cross sections are strongly enhanced \cite{Cardona:2010a,ivchenko05a,Guo:2015a}, as documented already in the pioneering experiments in GaAs based structures \cite{Miller:1986a,Cerdeira:1986a}. A model description of double resonant Raman processes for 1 and 2-photon excitation experiments is given in the supplement \cite{suppl}. \\
\indent Our next target is to analyse which phonon modes can be combined to bridge the 140~meV energy difference between the {$2s/2p$} and {$1s$} exciton states, respecting {the wavevector} conservation and symmetry (see supplement for details \cite{suppl}). The WSe$_2$ monolayer has $D_{3h}$ point symmetry and $P\bar 6m2$ (187 or $D_{3h}^1$) space group. The analysis of Refs.~\cite{Jimenez:1991a,Ataca:2011a,Luo:2013a,Ribeiro:2014a} allows to associate 9 phonon modes at the $\Gamma$ point (center of the Brillouin zone) with the following representations of $D_{3h}$ point group: $A_1' +2 A_2'' + 2E' + E''$. To be Raman-active, the phonon modes should transform as quadratic combinations of coordinates, i.e. according to the irreducible representations $A_1'$, $E'$, and $E''$~\cite{suppl}. Theoretical calculations of phonon energies in WSe$_2$ monolayers are presented in Refs.~\cite{Sahin:2013a,Luo:2013a} and peaks in Raman spectra are identified in Refs.~\cite{Zhang:2015a,Zhao:2013a}, with strong signals with well defined peaks attributed to the $ A_1'(\Gamma)$, $E'(\Gamma)$ and $3LA(\mathrm M)$ modes. The combinations of the prominent phonons $A_1'(\Gamma) \otimes 2E'(\Gamma)\otimes 3LA(\mathrm M)$, or $2A_1'(\Gamma) \otimes E'(\Gamma)\otimes 3LA(\mathrm M)$, or $3A_1'(\Gamma) \otimes 3LA(\mathrm M)$, or $3E'(\Gamma)\otimes 3LA(\mathrm M)$~\cite{note:3LA} have an energy of $\hbar\omega \approx 140$~meV, see Table~\ref{tab:freq}. This is very close to the energy separation between $2s/2p$ and $1s$ excitons in ML WSe$_2$ and can account for double resonant Raman scattering~\cite{suppl}. These combinations contain the modes transforming both according to $A_1'$ and $E'$ and, hence, are Raman-active in $\bar z(xx) z$ and in $\bar z(xy)z$ configurations, i.e. both in co- and cross-linear polarizations. 
Analysis shows that these multiphonon modes are consistent with both resonant Raman transitions from $2s$ to $1s$ and from $2p$ to $1s$ exciton states observed, respectively, in one-photon and two-photon excitation experiments~\cite{suppl}.
\begin{table}
\begin{center}
\caption{Selected phonon energies in WSe$_2$, after Ref.~\cite{Zhang:2015a}}
\begin{tabular}{c|c|c}  
\hline
Mode  & Raman shift [cm$^{-1}$] & Energy [meV] \\
\hline
$E'(\Gamma)$ & 248 & 30.73\\
$A'_1(\Gamma)$ & 250 & 30.98\\
$3LA(M)$ & 394 & 48.82\\
\hline
\end{tabular}\label{tab:freq}
\end{center}
\end{table} 

\begin{figure}[h]
\includegraphics[width=0.45\textwidth]{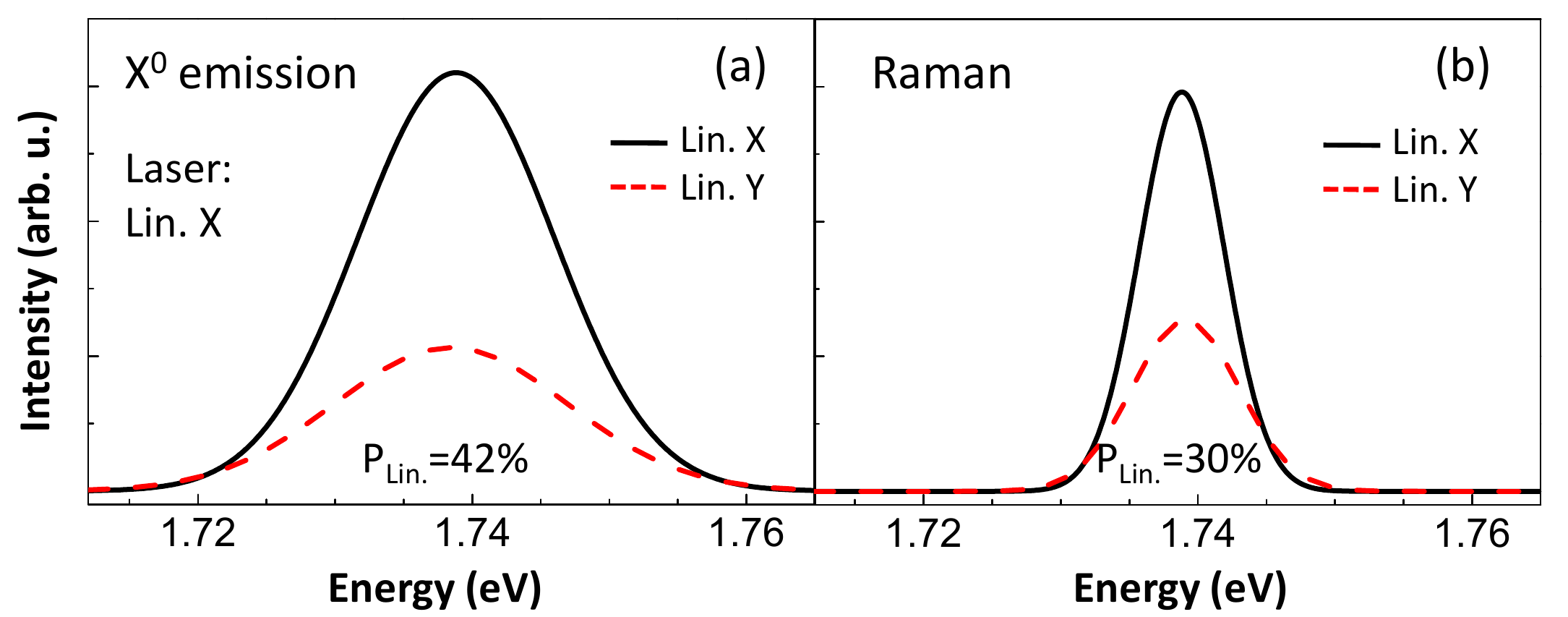}
\caption{\label{fig:fig3} \textbf{(a)} Following linearly polarized laser excitation 140~meV above the X$^0$ PL emission energy, the PL is strongly polarized. \textbf{(b)} The narrower double Raman resonance is also strongly linearly polarized. The sum of the graphs shown in (a) and (b) is a least square fit for the global emission with Gaussian line-shapes. We plot them here in different panels for clarity.
}
\end{figure}
\indent Our detailed analysis of double resonant Raman scattering allows to distinguish between the different contributions to the ML WSe$_2$ emission following excitation with linearly polarized laser light. As we can spectrally separate the X$^0$ PL from the Raman signal, we can analyze the polarization of each contribution individually, as shown in Fig.~\ref{fig:fig3}. For a laser energy 140~meV above the X$^0$ PL energy we observe a maximum of the global emission intensity, but also for each of the two components taken individually. For the X$^0$ emission, we observe a linear polarisation degree of 42\%, corresponding to strong valley coherence i.e. exciton alignment. From this time integrated measurement we can infer that the coherence is maintained during the PL emission time that we have determined to be in the ps range \cite{Wang:2014b}. For the Raman signal, we observe a linear polarization degree of 30\%. The depolarization of Raman line is qualitatively in agreement with the selection rules for the relevant modes, see above. 
In our experiment we successfully separate PL and Raman contribution to the signal, an important distinction for valley coherence control schemes. For other semiconductors this separation can be achieved in time-resolved experiments, as the scattered light is only present during the laser pulse whereas the PL signal persists for longer. This is not possible here as the PL signal decays within ps, i.e. during the excitation pulse of our ps laser, here experiments with fs time resolution would be helpful. Although the same physical processes analysed in this work will be applicable to experiments with femto-second laser pulses, in the spectral domain the Raman signal will appear much broader than the PL linewidth and will not be separable. \\
\indent \textit{Conclusions.---}
We analyze in detail the neutral exciton emission ($1s$) and the scattered light detected when exciting the $2s$ (or $2p$) exciton state with a linearly polarized laser. Tuning the laser in resonance with the  $2s$ (or $2p$) state has two effects: First, we observe double resonant Raman scattering. The incoming photons are resonant with the  excited exciton state, the outgoing photons with the  $1s$ exciton state. Second, we observe strong absorption, followed by efficient energy relaxation via emission of several phonons, as for example the combination $A_1'(\Gamma) \otimes 2E'(\Gamma)\otimes 3LA(\mathrm M)$. As a result the $1s$ exciton emission is strongly enhanced as compared to excitation with other laser energies. The PL emission is linearly polarized above 40\%, corresponding to very efficient valley coherence generation following the resonant exciton generation. Importantly, we are able to clearly separate the PL emission from the spectrally narrow Raman signal as we vary the laser excitation energy, which shifts the Raman feature.The multi-phonon resonance between the excited exciton states might be one of the reasons why efficient valley coherence can be generated in ML WSe$_2$, contrary to MoS$_2$ or MoSe$_2$, where this resonance might not occur. In particular we do not observe the double resonant Raman feature in PLE experiments on high quality MoSe$_2$ MLs around the $2p$ resonance \cite{Wang:2015c}. Further theoretical work is needed to reveal the nature (perturbative or strongly coupled) of the exciton-phonon interaction \cite{Verzelen:2002a}. Also the influence of magnetic fields applied perpendicular to the sample plane will allow to study separately the evolution of the polarization of the Raman signal and the valley coherence of the neutral exciton \cite{Aivazian:2015a,Wang:2015d}. The exact energy spacing and symmetry of the exciton states that govern the optical properties of ML TMDCs are still under debate with scattered values in the literature \cite{Cheiwchanchamnangij:2012a,Komsa:2012a,Song:2013a,He:2014a,Ugeda:2014a,Chernikov:2014a,Ye:2014a,Wang:2015b,Klots:2014a,Zhu:2015b,Hanbicki:2015a}. The demonstration of double resonant Raman scattering mediated by excitons introduces a sensitive tool to probing this material.   \\
\indent \textit{Acknowledgements.---} We acknowledge funding from ERC Grant No. 306719, ANR MoS2ValleyControl. M.M.G. is grateful to Labex NEXT for an invited Professorship, RFBR, RF President grant MD-5726.2015.2, and the Dynasty Foundation.

\newpage

\section*{SUPPLEMENTARY INFORMATION}
\section{Additional experimental data: One photon photoluminescence spectra}

\begin{figure}[h]
\includegraphics[width=0.9\linewidth]{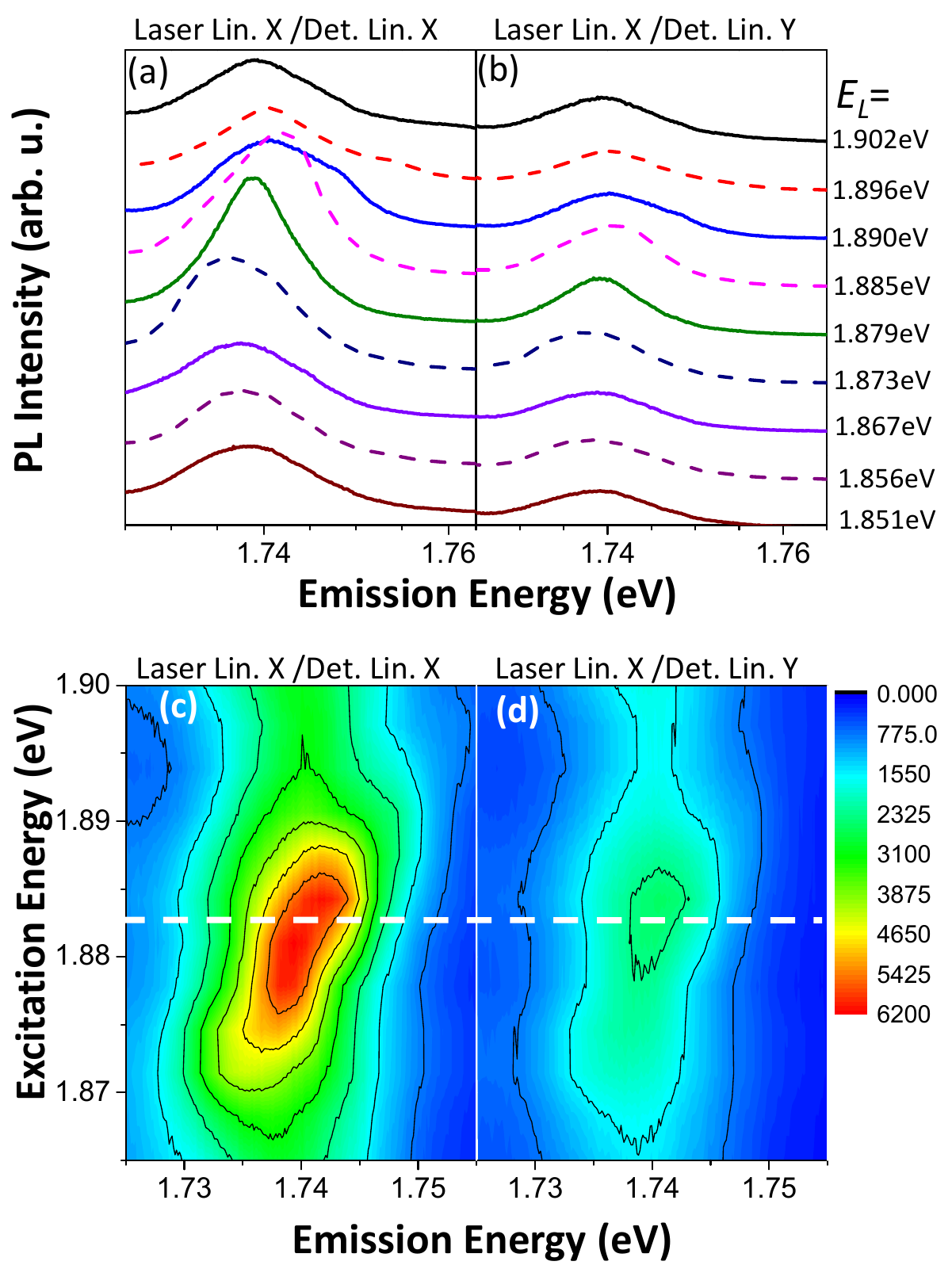}
\caption{\textbf{1-photon excitation (a)} emission at the X$^0$ energy (1s exciton) for different laser energies $E_L$ close to the 2s exciton resonance. Laser polarization linear X, detection linear X (co-polarized). \textbf{(b)} same as (a) but for detection  linear Y (cross-polarized). \textbf{(c)} same data as (a) but presented as a contour plot to show change in PL intensity and shape as a function of the excitation laser energy $E_L$. \textbf{(d)} same data as (b), same color scale as (c). Polarization configurations are marked on top of the panels.}\label{fig:supp:1pple}
\end{figure}

Figure~\ref{fig:supp:1pple} shows the 1-photon photoluminescence (PL) spectra measured at the conditions similar to those presented in the main text for two-photon PL. For these measurements the excitation laser energy was scanned in the vicinity of $2s$ exciton energy, $1.89$ eV. The change of shape of the PL emission as well as the variation of the position of the central energy with the energy of excitation laser is clearly seen.

\section{Symmetry of phonon modes in WSe$_2$ and selection rules for Raman process}\label{sec:symm}

\begin{figure}[h]
\includegraphics[width=\linewidth]{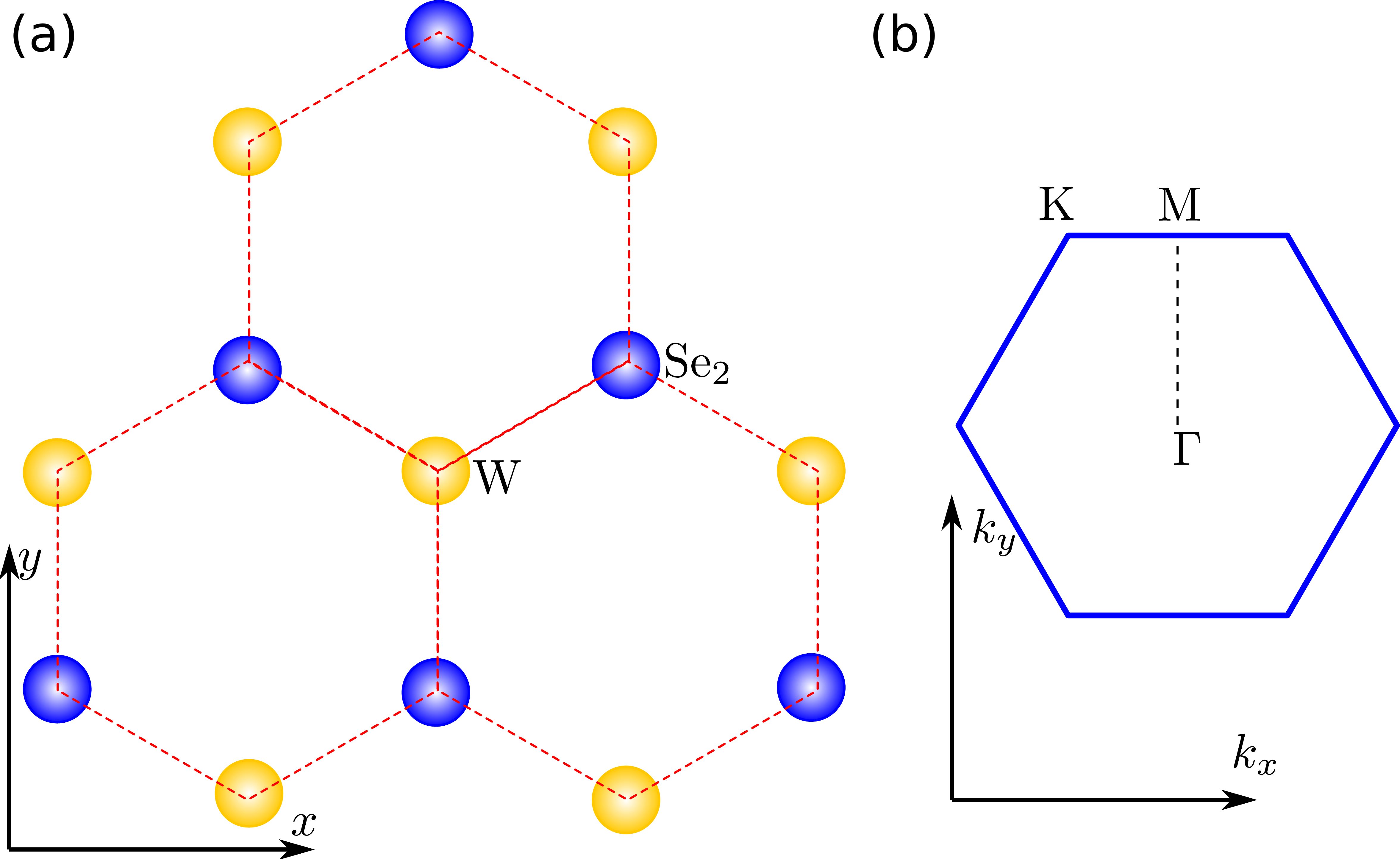}
\caption{\textbf{(a)} Lattice of WSe$_2$ projected on the Tungsten plane. \textbf{(b)} Brilloin zone of WSe$_2$.}\label{fig:latt}
\end{figure}

WSe$_2$ monolayer has $D_{3h}$ point symmetry and $P\bar 6m2$ (187 or $D_{3h}^1$) space group~\cite{PhysRevB.90.115438}. There are three atoms in the unit cell resulting in 9 phonon modes. The detailed symmetry analysis of the modes is presented in Refs.~\cite{PhysRevB.44.3955,doi:10.1021/jp205116x,PhysRevB.84.155413,PhysRevB.88.195313,PhysRevB.90.115438}. The symmetry of the phonon modes at the center of Brilloin zone can be established using correlation method. We choose a center of point transformations in the position of W atom. The symmetry of W site is, hence, $D_{3h}$, while the symmetry of the Se sites is $C_{3v}$, see Fig.~\ref{fig:latt}(a). In $D_{3h}$ point group the displacements transform according to $E'$ representation ($x,y$) and $A_2''$ representation ($z$), see Table~\ref{tab:repr}. In $C_{3v}$ point group the displacements transform according to $E$ ($x,y$) and $A_1$ ($z$). Making use of the {correlation table for $D_{3h}$ point group}~\cite{corr:tab} we obtain the following representations of the phonon modes~\cite{PhysRevB.44.3955,doi:10.1021/jp205116x,PhysRevB.88.195313,phonons:2d}:
\begin{equation}
\label{phonons:Gamma}
A_2'' + E' + E' + E'' + A_1' +A_2'' = 2 A_2'' + 2E' + E'' + A_1'.
\end{equation}
The acoustic modes at the $\Gamma$ point transform according to $A_2''$ (non-degenerate) and $E'$ (two-fold degenerate at the $\Gamma$ point), the remaining modes are optical.

We recall that the Raman scattering can be described as phonon-induced fluctuations of polarizability tensor $\chi$ which links the linear dielectric polarization of the medium $\bm P$ and electric field $\bm E$ as
\begin{equation}
\label{chi}
P_i = \chi_{ij} E_j,
\end{equation}
where $i$ and $j$ denote Cartesian components. For a single-phonon process the fluctuating part of $\delta \chi_{ij}$ is proportional to $Q^{(\alpha)}\exp{(\mp \mathrm i \Omega_\alpha t)}$, where $\alpha$ enumerates phonon modes (including their polarization and wavevector), $Q^{(\alpha)}$ are the generalized phonon coordinates, $\Omega_\alpha$ are their frequencies, and $\mp$ signs correspond to the absorption/emission of a phonon, respectively. Analogously, for multiphonon processes involving $\alpha,\beta,\ldots, \delta$ modes, we have 
\begin{equation}
\label{chi:fluct}
\delta \chi_{ij} \propto Q^{(\alpha)}Q^{(\beta)}\ldots Q^{(\delta)}\exp{[- \mathrm i (\pm\Omega_\alpha \pm\Omega_\beta \pm \ldots \pm\Omega_\delta) t]},
\end{equation}
giving rise to the combination frequencies in Raman spectra. To be Raman-active, the product $Q^{(\alpha)}Q^{(\beta)}\ldots Q^{(\delta)}$ should transform like products of coordinates $r_ir_j$, where $\bm r = (x,y,z)$.
Hence, the Raman-active modes in WSe$_2$ transform as  quadratic combinations of coordinates, i.e. according $A_1'$, $E'$, and $E''$.  Hence, the only Raman-inactive mode is $A_2''$ but the combined modes, e.g., $2A_2''$ are also active ($A_2''\times A_2'' = A_1'$). Note, that the modes with $E''$ symmetry are Raman-active at the oblique incidence only. 

\begin{table}[h]
\begin{center}
\caption{Representations of $D_{3h}$ point symmetry group}
\begin{tabular}{c|c}  
\hline
Representation  & Basic functions  \\
\hline
$A_1'$ & $x^2+y^2$; $z^2$\\
$A_2'$ & $J_z$\\
$E'$ & $(x,y)$; $(x^2-y^2,-2xy)$\\
$A_1''$ & $zJ_z$\\
$A_2''$ & $z$\\
$E''$ & $(J_x,J_y)$; $(xz,yz)$\\
\hline
\end{tabular}\label{tab:repr}
\end{center}
\end{table}

We are also interested in the symmetry of the phonons at the M point ($\bm K\parallel y$) of the Brillouin zone, Fig.~\ref{fig:latt}(b). The wavevector group is $C_{2v}$ ($\sigma_h$ becomes $\sigma_v$ and one of $\sigma_v$ planes, namely, containing $\Gamma$-M or $y$ direction, remains). In $C_{2v}$ point group displacements transform according to $A_1$ (invariant, $y$), $B_1$ ($z$) and $B_2$ ($x$) representations. Particularly, the longitudinal acoustic mode at the M-point, $LA(\mathrm M)$, transforms according to $A_1$ in $C_{2v}$ (note, that $E' = A_1+B_2$).

The multiphonon modes $A_1'(\Gamma) \otimes 2E'(\Gamma)\otimes 3LA(\mathrm M)$, or $2A_1'(\Gamma) \otimes E'(\Gamma)\otimes 3LA(\mathrm M)$, or $3A_1'(\Gamma) \otimes 3LA(\mathrm M)$, or $3E'(\Gamma)\otimes 3LA(\mathrm M)$, involve combinations of optical phonons $A_1'(\Gamma)$, $E'(\Gamma)$ at the Brillouin zone center and a combination of three longitudinal acoustic modes at M-point with zero net wavevector. They contain several $A_1'$ and $E'$ representations. Hence, these combinations are Raman-active in $\bar z(xx) z$ configuration and in $\bar z(xy)z$ configurations. Note, that due to the fact that these multiphonon states contain $A_1'$ and $E'$ representations they can result in the transfer of excitation from $2p$ exciton state (representation $E'\times E'=A_1'+A_2'+E'$, where the first representation in the product denotes the symmetry of the exciton Bloch function and the second one denotes the symmetry of the envelope) to $1s$ exciton state (representation $E'\times A'=E'$).

\section{Model of double resonant Raman scattering}

To illustrate the enhancement of the Raman scattering we model $2s$ and $1s$ excitonic states in WSe$_2$ as in-plane dipoles with the polarization $\bm P_{2s}$ and $\bm P_{1s}$ obeying harmonic oscillator equations~\cite{ivchenko05a}
\begin{subequations}
\label{polarizations}
\begin{align}
\ddot{\bm P}_{2s} + \omega_{2s}^2 \bm P_{2s} &= g \bm E_0 \cos{\omega t},\label{2s}\\
\ddot{\bm P}_{1s} + \omega_{1s}^2 \bm P_{1s} &= \hat S(t) \bm P_{2s},\label{1s}
\end{align}
\end{subequations}
where $\omega_{2s}$ and $\omega_{1s}$ are the energies of the corresponding excitonic states, $\bm E_0 \cos{\omega t}$ is the electric field of the incident radiation, whose frequency $\omega$ is assumed to be close to $2s$ exciton resonance, $g$ is the coupling strength and the matrix $\bm S(t)$ describes multi-phonon-induced interaction of the oscillators. Its matrix elements in Cartesian basis are $S_{ij} =S_{ij}^{(0)} \cos{\Omega t}$, where $\Omega$ is the frequency of the multi-phonon mode~\footnote{An arbitrary initial phase is not relevant here.} and the prefactors $S_{ij}^{(0)}$ contain contributions from virtual transitions involving all relevant phonon modes. Their calculation is highly complex since it involves high order perturbation theory and is beyond the scope of the present work~\footnote{We stress that unlike the multiphonon cascades observed in GaSe, where the intermediate states are real~\cite{Cardona:2010a}, here the intermediate states are all virtual and the cascade model~\cite{PhysRevLett.26.1241} is not applicable.}.
Equation~\eqref{1s} clearly shows that only (multi-)phonon modes, which transform as products of coordinates, $x^2$, $y^2$, $xy$ or their combinations (that are $A_1'$ and $E'$) can couple $2s$ and $1s$ excitons.

The solution of set of Eqs.~\eqref{polarizations} yields the polarization of the $1s$ exciton. It contains the contributions oscillating as $\cos{[(\omega \pm \Omega)t]}$ with the amplitudes
\begin{equation}
\label{ampl}
P_{1s,i} = \sum_{\pm,j} \frac{S_{ij}^{(0)}gE_{0,j}\cos{[(\omega \pm \Omega)t]}}{(\omega_{2s}^2 - \omega^2) [\omega_{1s}^2 - (\omega \pm \Omega)^2]}.
\end{equation}
Equation~\eqref{ampl} shows that the polarization of $1s$ exciton oscillates with the frequencies $\omega \pm \Omega$ which corresponds to the anti-Stokes and Stokes peaks~\footnote{The relative intensities of the Stokes and anti-Stokes components depend on the temperature, since Stokes process involves phonon emission, while anti-Stokes involves phonon absorption. In the oscillator model outlined here this effect can be included by quantum-mechanical treatment of correlators of $\hat S$ elements and evaluation of scattered light intensity.}. The denominators in Eq.~\eqref{ampl} clearly demonstrate the resonant enhancement of the Raman scattering~\footnote{Allowance for exciton damping suppressed divergences in Eq.~\eqref{ampl}.}. The double resonance occurs if incident radiation frequency is close to $\omega_{2s}$ and, additionally, if $\omega_{1s}$ is close to $\omega - \Omega$ (or $\omega+\Omega$ for the anti-Stokes scattering) and results in drastic enhancement of $\bm P_{1s}$ and, correspondingly, of the scattered light intensity.

Similar analysis demonstrates that the double resonance Raman scattering is possible under the conditions of two-photon excitation, where the $2p$ exciton resonance is addressed. Note, that in WSe$_2$ the $2s$ and $2p$ excitons are very close in energy~\cite{Wang:2015b}. The $2p$ exciton state can be conveniently modelled in the coupled-oscillator picture as a quadrupolar-like mode $U_{2p,ij}$, where one of the Cartesian subscripts ($i$) denotes the Bloch function of exciton, representation $E'$, and the other one  ($j$) denotes the envelope function ($p_x$-shell or $p_y$-shell like, representation $E'$) for $2p$ state. The set of Eqs.~\eqref{polarizations} is replaced by
\begin{subequations}
\label{2photon}
\begin{align}
\ddot{U}_{2p,ij} + \omega_{2p}^2 U_{2p,ij} &= \sum_{kl} g'_{ij,kl} E_{0,k}E_{0,l} \cos{2\omega t},\label{2p}\\
\ddot{P}_{1s,i} + \omega_{1s}^2  P_{1s,i} &=  \sum_{jl} S'_{i,jl}(t) U_{2p,jl},\label{1s:2p}
\end{align}
\end{subequations}
where the tensor $g'_{ij}$ describes the selection rules at two photon absorption and components $S'_{i,jl}\propto \cos{\Omega t}$ describe contributions for multiphonon processes. For in-plane oscillators where $i,j,l=x$ or $y$ the phonons of $E'$ and $A_1'$ symmetry can provide such a coupling, i.e. the same modes which provide $2s$ to $1s$ transfer. The solution of Eqs.~\eqref{2photon} can be presented in a form, similar to Eq.~\eqref{ampl}:
\begin{equation}
\label{ampl:2phot}
P_{1s,i} = \sum_{\pm,jlkm} \frac{S_{i,jl}^{(0)}g'_{jl,km} E_{0,k}E_{0,m}\cos{[(2\omega \pm \Omega)t]}}{(\omega_{2p}^2 - 4\omega^2) [\omega_{1s}^2 - (2\omega \pm \Omega)^2]}.
\end{equation}
Here the double resonance occurs when the twice the photon energy $2\omega$ approaches $\omega_{2p}$, the frequency of the $2p$ state, and the frequency of the secondary light, $2\omega \pm \Omega$, approaches the frequency of the ground exciton state, $1s$.

The experimental data presented in the main text and in Fig.~\ref{fig:supp:1pple} clearly show that the double resonance Raman signal is superimposed on the PL. To analyse the origin of these two contributions to the emission and demonstrate that such a superposition is indeed consistent with the model presented above, let us consider the correlation function of the source terms for $\bm P_{1s}$ in Eqs.~\eqref{1s}, \eqref{1s:2p}. To be specific, we consider the case of 1-photon excitation, Eqs.~\eqref{polarizations}, and introduce the correlation functions as
\begin{equation}
\label{corr}
\Xi_{ij} =
\sum_{kl}\langle \cos{[\omega_{2s}(t+\tau)]} S_{ik}(t+\tau) P_k(t+\tau) S_{jl}(t)P_l(t) \rangle,
\end{equation}
where the averaging is performed over the time $t$ at a fixed delay $\tau$; the factor $\cos{(\omega_{2s}\tau)}$ is included into Eq.~\eqref{corr} to exclude the oscillations of $\Xi_{ij}(\tau)$ at the optical frequency. The correlation functions $\Xi_{ij}$ quantify the extent to which the driving force in the right hand side of  Eq.~\eqref{1s} is harmonic. Particularly, $\Xi_{ij}$ can be presented as a sum of contribution of all possible multiphonon modes. The terms of interest for us oscillate at a frequency $\Omega\approx \omega_{2s} - \omega_{1s}$, which provide the double resonance effect:
\begin{equation}
\label{decomp}
\Xi_{ij}(\tau) = \sum_{\alpha} \Xi_{ij}^{0} \cos{(\Omega \tau)} \exp{(-\gamma_\alpha \tau)}.
\end{equation}
Here the subscript $\alpha$ runs through all relevant multiphonon combinations, see Sec.~\ref{sec:symm}, and we introduced the mode damping rates $\gamma_\alpha$, small differences of the multiphonon combination frequencies are disregarded, possible effect of phonon modes dispersion is included into the damping rates $\gamma_\alpha$. 

 For the modes with $\Omega/\gamma_\alpha \gg 1$ the correlation function oscillates with time and weakly decays. These modes provide double resonant Raman scattering as the induced polarization $\bm P_{1s}$ oscillates with time as $\cos{[(\omega_{2p} - \Omega)t]}$ (or as $\cos{[(\omega_{2p} + \Omega)t]}$ for anti-Stokes scattering). Other modes, where $\Omega/\gamma_\alpha \lesssim 1$, do not act as a harmonic driving force for the $1s$-exciton polarization, since their correlation function decays at a time scale, $\gamma_{\alpha}^{-1}$, being short as compared with the oscillation period. These modes act as random (Langevin) forces and result in the incoherent population of $1s$ state, giving rise to oscillations of $\bm P_{1s}$ with the $1s$ exciton frequency $\omega_{1s}$ and yielding the photoluminescence superimposed on the Raman signal.


\begin{thebibliography}{59}
\bibitem[{pho()}]{phononbroad}
\bibinfo{note}{Measuring 5~meV and not 3~meV as given by the laser linewidth
  might be due to broadening by phonon dispersion and damping.}
\bibitem[{not({\natexlab{b}})}]{note:3LA}
\bibinfo{note}{The combination of three acoustic phonons at M point of the
  Brillouin zone has zero total wavevector and is Raman active~\cite{suppl}.}
\end{thebibliography}

\end{document}